\documentclass{ws-procs11x85}
\usepackage{balance} 

\newcommand{\met}{\hbox{E\kern-0.5em\lower-0.1ex\hbox{/}}_T}

\newcommand{\apj} {ApJ}

\newcommand{\prl} {Phys. Rev. Letts.}
\newcommand{\pra} {Phys. Rev. A.}
\begin{document}

\twocolumn[
\title{Long Term Evolution of Magnetic Turbulence in Relativistic Collisionless Shocks}

\author{Philip Chang, Anatoly Spitkovsky, and Jonathan Arons}
\address{Department of Astronomy, Campbell Hall, University
  of California, Berkeley, CA 94720\\
Department of Astrophysical Sciences, Peyton Hall,
Princeton University, Princeton, NJ 08544\\
Email: pchang@astro.berkeley.edu, anatoly@astro.princeton.edu, arons@astro.berkeley.edu}


\begin{abstract}
 We study the long term evolution of magnetic fields generated by an
 initially unmagnetized collisionless relativistic $e^+e^-$ shock.
 Our 2D particle-in-cell numerical simulations show that downstream of
 such a Weibel-mediated shock, particle distributions are
 approximately isotropic, relativistic Maxwellians, and the magnetic
 turbulence is highly intermittent spatially, nonpropagating, and
 decaying. Using linear kinetic theory, we find a simple analytic form
 for these damping rates.  Our theory predicts that overall magnetic
 energy decays like $(\omega_p t)^{-q}$ with $q \sim 1$, which
 compares favorably with simulations, but predicts overly rapid
 damping of short wavelength modes.  Magnetic trapping of particles
 within the magnetic structures may be the origin of this discrepancy.
 We conclude that initially unmagnetized relativistic shocks in
 electron-positron plasmas are unable to form persistent downstream
 magnetic fields.  These results put interesting constraints on
 synchrotron models for the prompt and afterglow emission from GRBs.
\end{abstract}

\vskip12pt  
]

\bodymatter

\section{Introduction}


The prompt emission and afterglows of gamma-ray bursts (GRBs) may be
manifestations of ultrarelativistic shock waves.  These shock waves
may be mediated via the relativistic form of Weibel instability
(Weibel 1959; Yoon and Davidson 1987; Medvedev and Loeb 1999; Gruzinov
and Waxman 1999).  The free energy from strong plasma anisotropy in
the shock transition layer generates strong magnetic fields (with
strengths comparable to the available free energy).  However, these
fields have very small spatial scales, i.e., the order of the plasma
skin depth, $c/\omega_p$, where $\omega_p$ is the plasma frequency.
These initially small-scale B-fields must survive for tens of
thousands to millions of inverse plasma periods to serve as this
source of the magnetization for synchrotron models of burst emission
and afterglows (Gruzinov \& Waxman 1999; Piran 2005ab; Katz, Keshet,
\& Waxman 2007). Whether or not these field can is an open question.


Numerical and analytic studies (Kazimura {\it et al.} 1998; Silva {\it
  et al.} (2003); Frederiksen {\it et al.}  2004; Medvedev {\it et
  al.} 2005; Hededal {\it et al.}  2005; Nishikawa {\it et al.}  2003,
2005; Spitkovsky this proceedings) have elucidated the basic
physics. The instability initially forms filaments of electric current
and $B$ fields, which then merge to {\it inverse cascade} magnetic
energy to larger scales, but only in the {\it foreshock} region.  When
the B-fields reach the magnetic trapping limit (Davidson {\it et al.}
1972; Kato 2005; also see Milosavljevic, Nakar, \& Spitkovsky 2006;
Milosavljevic \& Nakar 2006a), particle orbits become chaotic,
disorganizing the filaments.  The disorganized magnetic fluctuations
scatter their supporting particles, which isotropizes and thermalizes
the flow, within tens to hundreds of skin depths (Spitkovsky 2005).
The magnetic energy peaks in this layer at $\sim$10-20\% of the bulk
plasma flow energy.

However, present simulations have not deeply followed the flow into
the downstream region to explore the long term behavior of these
B-fields.  Thus, the question of the structure and long-term survival
of the B-fields remains open (see for instance, Gruzinov \& Waxman
1999; Gruzinov 2001ab; Medvedev {\it et al.} 2005).

In this proceeding, we discuss recent work (Chang, Spitkovsky, and
Arons 2008; hereafter CSA08) which shows that this magnetic energy must
rapidly decay in the downstream medium.  We first describe the basic
features of the downstream plasma from our numerical simulations.
We then calculate the evolution (decay) of the downstream plasma using
Vlasov linear response theory and then compare this evolution with
simulations.  While linear theory does reasonably well in estimating
the decay rate of the total magnetic energy, it overestimates the
damping rate of shorter wavelength modes.  We discuss this discrepancy
as a result of magnetic trapping. Finally, we summarize our results.

\section{Simulation Results}\label{sec:simulation}

Spitkovsky (2005, this proceedings) and Spitkovsky and Arons (in prep)
describe a series of 2D and 3D simulations of relativistic shock waves
in $e^+e^-$ plasmas.  These are Particle-in-Cell (PIC) simulations,
using the code {\it TRISTAN-MP}.  We simulate shocks by injecting cold
relativistic plasma particles at one end of a large domain that
reflect off a fixed conducting wall at the other end.  We use 2D boxes
as large as 50,000 x 2048 cells with up to $1.35\times 10^{10}$
particles to study these shocks. We refer the reader to CSA08 for
additional details.

\begin{figure}
\centerline{\psfig{file=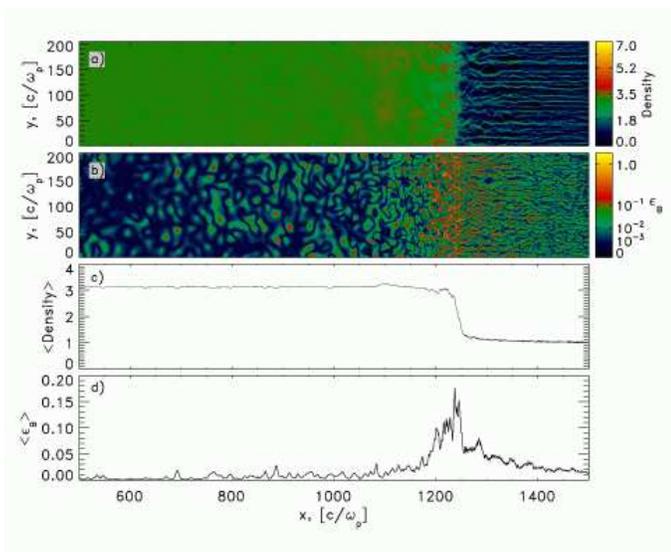,width=9cm}}
\caption{Snapshot of a region from a large 2D relativistic shock
  simulation in the downstream frame.  a) Density structure,
  normalized in upstream units, in the simulation plane. b) Magnetic
  energy, normalized in terms of upstream energy of the incoming flow:
  $\epsilon_B=B^2/4 \pi \gamma_1 n_1 m c^2$.  A power law scaling,
  $\epsilon_B^{1/4}$, was applied to stretch the color table to show
  weak field regions and is reflected in the colorbar.  c) Plasma
  density averaged in the transverse direction as a function of the
  distance along the flow.  d) Magnetic energy density averaged in the
  transverse direction, as a function of distance along the flow.}
\label{fig:shock}
\end{figure}

Figure \ref{fig:shock} shows the snapshots of density and magnetic
energy from a typical 2D simulation.  Coordinates are in units of the
upstream skin depth, $c/\omega_{p1}$. In the simulation shown, the
upstream flow moves to the left with $\gamma_1 = 15$.

Our simulations are large enough to permit the complete development of
the shock and show the main features of contemporary collisionless
shock simulations.  For instance, we see the factor of $\approx 3.13$
increase in density between the upstream and the downstream (Fig.
\ref{fig:shock}c), which is the expected compression factor (Gallant
{\it et al.}  1992; Spitkovsky and Arons, in prep).  Current filaments
show up as an enhancement in the plasma density and magnetic energy
density in the foreshock (Fig.  \ref{fig:shock}ab). The scale of the
filaments grows towards the shock through merging.

At the shock transition layer, the filaments disorganize and become
clumps of magnetic energy.  These magnetic clumps lose intensity the
further downstream they are from the shock (Figure \ref{fig:shock}b).
We also find that the particle distribution function changes to an
isotropic (in the downstream rest frame) thermal population, i.e., the
difference between the perpendicular and parallel momentum is
extremely small, $<1\%$.  As Figures \ref{fig:shock} b and d suggest,
the magnetic fields decay in the downstream region of the shock.

\section{Downstream Evolution of Magnetic Turbulence}\label{sec:theory}

We now attempt to analytically understand this decay of magnetic
turbulence. The simulations show that the downstream plasma is
isotropic and the downstream particle distribution function is well
described by a relativistic Maxwellian. We will also {\it assume} the
downstream field amplitudes are so small that particle orbits are
almost straight lines.  

We begin by deriving the linear plasma response is determined by the
plasma susceptibility, $\chi$ (Stix 1992).  We evaluate the susceptibilty for
distribution functions that are isotropic in two and three
dimensions. Details can be found in CSA08.  We set $\omega_r = 0$,
because of the non-propagating nature of the magnetic clumps, which we
infer from the simulations.  In the long wavelength limit, i.e., $k
\ll \omega_p/c$, where $k$ is the wavenumber, we find:
\begin{equation}\label{eq:chi_limit}
4\pi\chi \approx \left\{\begin{array}{ll} 
i \frac {\omega_{p}^2}{|k|c\omega} & \textrm{2D} \\
i \frac {\pi} 4 \frac {\omega_{p}^2}{|k|c\omega} & \textrm{3D}
\end{array}\right. . 
\end{equation}
Note that the 2D and 3D results only vary by a numerical factor.  Long
wavelength modes have the same qualitative behavior in two and three
dimensions.

The plasma susceptibility (eq.[\ref{eq:chi_limit}]) can be utilized to
calculate the evolution of an initial field of fluctuations.  We refer
the interested reader to CSA08 for details, but the
result is
\begin{equation}\label{eq:damping}
\frac {d|\delta B_k|^2}{dt} = -2\gamma_k |\delta B_k|^2,
\end{equation}
where $\gamma_k = \left(kc\right)^2\omega^{-1}
\Im\left(4\pi\chi\right)^{-1} $.  The asymptotic forms of $\chi$ from
equation (\ref{eq:chi_limit}) for 2D and 3D gives
\begin{equation}\label{eq:gamma}
\gamma_{k} = \left\{\begin{array}{ll} \frac {|kc|^3} {\omega_{p}^2} & \textrm{2D} 
\\ \frac {4} {\pi} \frac{|kc|^3} {\omega_{p}^2}& \textrm{3D}
\end{array}\right. .
\end{equation}
Note the strong cubic $k$ dependence on the decay rate.  Short
wavelength modes rapidly damp, but longer wavelength modes can
persist.

\begin{figure}
\centerline{\psfig{file=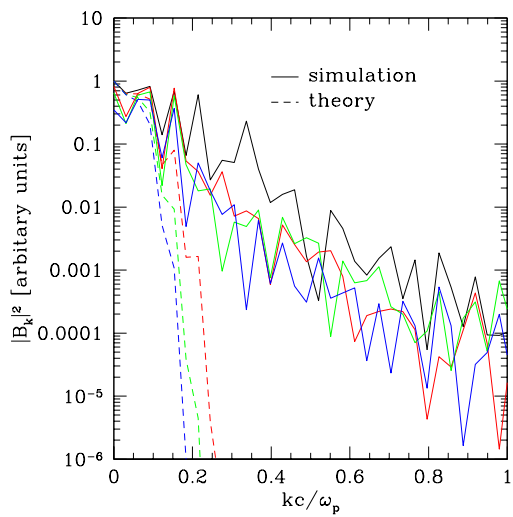,width=7cm}}
\caption{ Spectral evolution of magnetic field from the slice at $840
  c/\omega_p$. Initial field spectrum (black solid line) is plotted
  after $450\,\omega_p^{-1}$ (red), $900\,\omega_p^{-1}$ (green), and
  $1350\,\omega_p^{-1}$ (blue) based on simulation data. Dashed curves
  represent analytic evolution of the initial field.
\label{fig:fftsingle} }
\end{figure}

We now compare these expectations to the numerical simulations.  We
take the Fourier transform of $\delta B$ from our 2D numerical
simulations from a downstream region behind the shock front at $ x =
x_0$.  We evolve these spectra for 450 (red), 900 (green), and 1350
$\omega_p^{-1}$ (blue) using equation (\ref{eq:gamma}). We compare
this analytically evolved spectra to Fourier transformed snapshots
taken from our numerical simulations at these times in Figure
\ref{fig:fftsingle}, where $x_0=840 c/\omega_p$.  While theory and
simulation agree at very low wavenumber ($k_yc/\omega_p \lesssim
0.2$), theory overpredicts the cutoff in power at larger k.  The
discrepancy may be due to magnetic trapping (see \S\ref{sec:trap}).


\begin{figure}
\centerline{\psfig{file=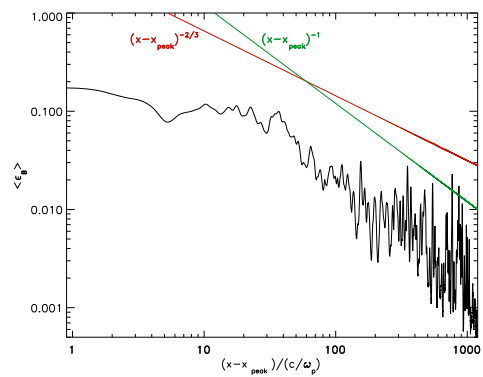,width=7cm}}
\caption{Magnetic energy density (in units of upstream kinetic energy)
  as a function of position downstream of the shock. A broken power
  law proportional $(x-x_{\rm peak})^{-2/3}$ fits well at early times,
  but a $(x-x_{\rm peak})^{-1}$ power law fits better at later times.
}
\label{fig:eb}
\end{figure}

Since total B-field energy is dominated by long wavelength modes, we
use equation (\ref{eq:damping}) to find a simple decay law for the
total B-field.  Again we refer the interested reader to CSA08 for
details, but if the initial spatial spectrum is a power law in wave
number $|\delta B_k|^2 \propto k^{2p}$, then the B-field should decay
like 
\begin{equation}
  \delta B^2 \propto t^{-2(p+1)/3}.
\end{equation}
For a shock moving at constant velocity, we have $x_{\rm peak} - x
\propto t$.  Hence $\delta B^2 \propto (x_{\rm peak} -
x)^{-2(p+1)/3}$.  Our numerical simulations are extremely suggestive
that the magnetic energy density follows the a $\delta B^2/8\pi
\propto t^{-2/3}$ decay expected for an initially flat magnetic
spectrum at early times ($p=0$), then steepening to a $t^{-1}$ decay
at later times ($p=1/2$) as shown in Figure \ref{fig:eb}.  We have
analyzed additional simulations with a large transverse spatial scale
and they suggest $p=0$.  This difference in the index of the decay law
expected from the theory and measured from the simulations may also be
due to magnetic trapping (see \S\ref{sec:trap}; see Gruzinov 2001b for
an alternate explanation and CSA08 for a rebuttal).

\section{Magnetic Trapping}\label{sec:trap}

While our simulations and theory are consistent with one another in
suggesting overall power law decay $t^{-q}$ with $q \sim 1$, they
disagree for short wavelengths. This discrepancy may arise from the
nonlinear effects of magnetic trapping.  Our simulations also show
that not all particles follow straight line trajectories that are
weakly perturbed, but some are partially trapped and strongly
deflected.  By following test particle orbits in our simulations, we
find that the Larmor radii of many of the test particles are of the
same order of the sizes of these clumps or smaller.  These strong
departures from weakly perturbed particle dynamics may be the cause of
the decreased damping at large wavenumber found in the simulations.
It may also modify the overall decay away from $t^{-2/3}$ decay law
expected from simple linear theory.



\section{Discussion}\label{sec:discussion}

We have studied the downstream evolution of magnetic turbulence in the
context of a collisionless $e^+e^-$ shock both analytically and
numerically.  Our simulations show that the downstream region consist
of nonpropagating magnetic clumps embedded in quasi-homogenous medium
where the background particle distribution function is an isotropic
Maxwellian.  In such a background, we showed that magnetic energy will
decay like $t^{-q}$ with $q\sim 1$.  However, linear theory
overpredicts the decay rates at short wavelengths compared to
simulations.  Magnetic trapping may play an role in resolving this
discrepancy.  Rapid field decay puts severe constraints on GRB
emission mechanism, but they may not be inconsistent with GRB
observations (see Pe'er \& Zhang 2006).  Finally, if ion-electron
collisionless shocks reach roughly equipartition with each other as
suggest by recent large scale simulations (Spitkovsky 2008), they
would reproduce the physics of the $e^{\pm}$ shock and their B-fields
would decay as well.


\section*{Acknowledgements}

We thank S. Cowley, D. Kocelski, M. Milosavljevic, A. Pe'er and E.
Quataert for useful discussions. 
P.C.
is supported by the Miller Institute for Basic Research. J.A. has
benefited from the support of NSF grant AST-0507813, NASA grant
NNG06GI08G, and DOE grant DE-FC02-06ER41453, all at UC Berkeley; by
the Department of Energy contract to the Stanford Linear Accelerator
Center no. DE-AC3-76SF00515; and by the taxpayers of California.
A.S. is pleased to acknowledge that the simulations reported on in
this paper were substantially performed at the TIGRESS high
performance computer center at Princeton University which is jointly
supported by the Princeton Institute for Computational Science and
Engineering and the Princeton University Office of Information
Technology.

\end{document}